\documentclass[%
  reprint,
  twocolumn,
 showpacs,
 showkeys,
 preprintnumbers,
 amsmath,amssymb,
 aps,
  prl,
  longbibliography,
 ]{revtex4-1}

\usepackage[dvipsnames]{xcolor}

\usepackage{mathptmx}

\usepackage{amssymb,amsthm,amsmath}

\usepackage{tikz}
\usetikzlibrary{calc,decorations.pathreplacing,decorations.markings,positioning,shapes}

\usepackage[breaklinks=true,colorlinks=true,anchorcolor=blue,citecolor=blue,filecolor=blue,menucolor=blue,pagecolor=blue,urlcolor=blue,linkcolor=blue]{hyperref}
\usepackage{graphicx}
\usepackage{url}

\begin{document}

\title{Classical predictions for intertwined quantum observables are contingent and thus inconclusive}

\author{Karl Svozil}
\email{svozil@tuwien.ac.at}
\homepage{http://tph.tuwien.ac.at/~svozil}

\affiliation{Institute for Theoretical Physics,
Vienna  University of Technology,
Wiedner Hauptstrasse 8-10/136,
1040 Vienna,  Austria}

\date{\today}

\begin{abstract}
Classical evaluations of configurations of intertwined quantum contexts induce relations, such as true-implies-false, true-implies-true~\cite{2018-minimalYIYS}, but also nonseparability among the input and output terminals. When combined, these exploitable configurations (aka gadgets) deliver the strongest form of classical value indefiniteness. However, the choice of the respective configuration among all such collections, and thus the relation of its terminals, remains arbitrary and cannot be motivated by some superselection principle inherent to quantum or classical physics.
\end{abstract}


\maketitle

\section{Quantum clouds as collections of intertwined contexts and their classical doubles}

Quantum logic, as conceived by von Neumann~\cite{v-neumann-49,v-neumann-55} and Birkhoff~\cite{birkhoff-36},
is about the formal/theoretical universe of potential empirical observable propositions,
as well as the algebraic relations and operations among them.
Every single one of these observables is considered operational ``subject to the experimenter's grace''
as its actual measurement reflects the experimenter's (subjective) choice to indeed measure one of these
potential observables, {\it versus} its (often continuity of) complementary ones.
(This choice is mostly supposed to be {\it ``ex machina''}; that is, outside of the reach of quantum theory,
and not subject to nesting~\cite{everett,wigner:mb,everett-1956}.)
Thereby, all the other, then counterfactual, observables remain in a ``dormant/hypothetical'' realm,
an idealist~\cite{berkeley,stace} ``limbo'' of sorts.

Even explorations allowing logical operations exclusively among simultaneously commeasurable observables~\cite{kochen2,kochen3}
and permitting partial value definiteness~\cite{2012-incomput-proofsCJ,2015-AnalyticKS}
(in the recursion-theoretic sense of Kleene~\cite{Kleene1936})
rely upon, and are thus valid relative to, such collections of counterfactuals.
Thereby the predictions/forecasts
derived not for a single such collection of observables
-- here sometimes referred to as cloud or gadget --
but for (finite) selections from the multitude of conceivable (finite) collections of observables
may be inconsistent.

One way to conceptualize the (nonclassical) performance of quantized systems is in terms of (black) boxes
with input and output terminals as interfaces~\cite{2018-minimalYIYS}.
Like zero-knowledge proofs~\cite{Quisquater1990} (a topic the late Specker became interested in)
they are supposed to certify that they act ``truly quantum mechanically''
while at the same time not allowing any inspection (e.g., duplication or opening)
other than their performance at the input-output
terminals.
Fulfillment/certification is usually obtained by the exhibition of certain features or signals usually not encountered by classical devices, among them complementarity
and classical value indefiniteness (mostly pronounced as contextuality).
For a great variety of such criteria see Table~\ref{2018-c-table1-conc-rel-info},
as well as the references cited therein, later.

However, the signals obtained from these boxes are far from plain.
Indeed in what follows we shall argue that,
depending on which hypothetical configurations of (necessarily complementary) ``intrinsic'' observables are
considered, any individual outcome can {\it ad hoc} be classically (re)interpreted as an indication of nonclassicality.
Yet the same outcome could also be in full conformity with a classical interpretation.
A combination of such classical models allows any statistical prediction at the terminals.
Moreover,
there does not seem to exist any convincing reason to choose one
of such configurations over another, thus giving rise to either contradictory or arbitrary {\it ad hoc} signal analysis.

Already Specker~\cite{specker-60} contemplated about generalized exotic behaviors even beyond quantum boxes,
whereby his criteria for ``weirdness'' were inspired by scholastic counterfactuals
(aka {\it Infuturabilien}).
And
the benign outcome of his fable was only made possible by the unmarried daughter's determined alas futile attempt to open
the ``wrong'' -- according to her father's strategy -- box; at which point he gave in, and marriage ensued.
Quantum boxes and as will be later argued quantum clouds are not dissimilar: because
of complementarity
and classical value indefiniteness (aka contextuality)
complete knowledge of the situation is impossible by any known physical means.
An immediate idealistic~\cite{berkeley,stace,Goldschmidt2017-idealism}
objection to the use of counterfactuals could
be that the presupposition of the sort of omni-realism required for a classical analysis of quantum boxes
cannot be operational~\cite{bridgman} and supported by quantum mechanics.
Indeed, the partial algebra approach of Kochen and Specker~\cite{kochen2,kochen3,kochen1}
disallows operations among complementary observables
whilst making heavy use of intertwined collections of complementary maximal operators (aka contexts).

However, even classical models based on set representable partition logics~\cite{svozil-2001-eua}
such as Moore's initial state identification problem~\cite{e-f-moore}
and also Wright's generalized urn model~\cite{wright:pent,wright}
mimic quantum complementarity to a certain degree -- indeed,
formally up to quantum logics with separable sets of two-valued states~\cite[Theorem~0, p.~67]{kochen1}.
Thereby nonseparability of quantum observables with respect to the set
of two-valued states (interpretable as classical truth assignments)
serves as a strict criterion for nonclassicality,
and also as a criterion against realizations by set-theoretical representable partition logics,
even if such two-valued states exist.

In what follows we shall further exploit counterfactual
configurations of contexts which are intertwined (this terminology is borrowed from Gleason~\cite{Gleason})
in one or more common observable(s).
Such counterfactual configurations of contexts will be called {\em clouds}.

Graph theoretically~\cite[Appendix]{Godsil-Newman-2008}
a context can be associated with a complete graph (aka clique).
Its vertices are identified with the elements of the context.
Adjacency is characterized by comeasurable exclusivity.
Clouds are represented by collections of complete graphs (aka clicks or contexts) intertwining at the respective vertices,
thereby leaving the edges unchanged.

In quantum mechanics, contexts are identified with orthonormal bases,
or equivalently with the maximal operators~\cite[\S~84, Theorem~1, p.~171]{halmos-vs}
which can be (nonuniquely) formed by nondegenerate sums containing all the one-dimensional orthogonal projection operators
associated with those respective bases.
Elementary propositions are formalized by vectors of these bases of $d$-dimensional Hilbert space,
or by the orthogonal projection operators associated with such vectors~\cite{birkhoff-36}.
Graph theoretically the vertices are represented by the basis vectors,
and adjacency stands for orthogonality of these vectors;
that is, the edges represent the (pairwise) orthogonality
relations between the vectors (vertices).
(Each vertex must be connected to all the other $d-1$ vertices in the respective context by an edge.)
Thereby the graph representing a cloud
has a faithful orthogonal representation~\cite{lovasz-89,Portillo-2015}
in terms of the elements of the bases representing the respective contexts.
The inverse problem of finding some faithful orthogonal representation of a given graph is still open.
A necessary condition for the existence of intertwines is that the dimensionality of the
vector space is higher than two because in fewer dimensions than three contexts are either identical or disjoint.

Orthogonality hypergraphs~\cite{greechie:71} are compact representations of clouds
which reveal their structural constituents by signifying contexts/cliques/bases:
every complete graph $K_d$  is replaced by a {\em single smooth curve} (usually a straight line)
containing distinguished points that represent the vertices.
Thereby,  the $d(d - 1)/2$ edges of any such complete graph $K_d$ are replaced with a single smooth curve.
All the vertices ``within'' this smooth curve represent the mutually orthogonal vectors forming a $d$-dimensional basis.

Clouds may have various model realizations and representations: a particular cloud may have
\begin{itemize}
\item[(i)] a quantum mechanical realization in terms of intertwining orthonormal bases, as mentioned earlier;
\item[(ii)] a pseudo-classical realization in terms of partition logic which in turn have automaton logic or generalized urn models;
\item[(iii)] a classical realization if there is only a single context involved;
\item[(iv)] none of the above (such as a tightly interlinked ``triangle'' configuration of three contexts with two vertices per context).
\end{itemize}
Suffice it to say that (i) does not imply (ii), and vice versa.
Case (iii) can be interpreted as
a subalgebra of all the other groups enumerated, as the cases (i) and (ii) are pastings of contexts or (Boolean) blocks~\cite{nav:91}.

\section{Enforcing classical two-valued states}

The commonly used method for exploring nonclassicality is to consider
configurations of type (i) with a quantum realization, upon which a classical interpretation, if it exists,
is ``enforced'' in terms of uniform classical truth and falsity allocations of the associated propositions.
Such value assignments
can be formalized by two-valued   states $s \in \{0,1\}$ or (classical truth) value assignments
which are additive and add up to one whenever the propositions are exclusive and within a single context.

The physical intuition behind this formalization is the observation that any $d$-dimensional context or maximal observable
can be interpreted as an array of detectors after a $d$-port beam splitter~\cite{rzbb}.
In an ideal experiment, only one detector clicks (associated with the proposition that the system is in the respective state),
whereas all the other $d-1$ detectors remain silent.

Such uniform classical interpretations are supposed to be context-independent;
that is, the value on intertwining observables which are common to two or more contexts is independent
of the context.
Besides context-independence of truth assignments at the intertwining observables various variants of such measures assume conditions of increasing strength:
\begin{itemize}
\item[(I)] The ``measures'' or value assignments employed in so-called ``contextuality inequalities''
merely assume that every proposition is either true or false,
regardless of the other propositions in that context which are simultaneously measurable~\cite{cabello:210401}.
This allows all possible $2^d$ possibilities of value assignments in a $d$-dimensional context with $d$ vertices,
thereby vastly expanding the multitude of possible value assignments.
With this expansion, all Kochen-Specker sets trivially allow value assignments.
\item[(II)] The prevalent assumption of two-valued states or value assignments, also used by Kochen and Specker~\cite{kochen1} as well as Pitowsky~\cite{pitowsky:218},
is that only a {\em single} one of the $d$ vertices within a $d$-dimensional context
is true, and all the others are false; therefore any isolated $d$-dimensional context can have only $d$ such standard two-valued value assignments.
\item[(III)] An even more restricted rule of value assignment abandons uniform definiteness
and supposes~\cite{2012-incomput-proofsCJ,PhysRevA.89.032109,2015-AnalyticKS} that,
if all $d-1$ but one vertex  in a $d$ dimensional context are false, the remaining one is true,
and if one vertex within a $d$-dimensional context is true, all remaining  $d-1$ vertices are false.
This latter value assignments allow for {\em partial} functions which can be undefined.
\end{itemize}
Type (III) implies
type (II) which in turn implies type (I) value assignments.

A set $S$ of two-valued states on a graph $G$ is~\cite{svozil-tkadlec,tkadlec-96}:
\begin{itemize}
\item[(u)] {\em unital}, if for every $x\in G$ there is a two-valued state $s\in S$
such that $s(x) = 1$;
\item[(s)] {\em separating}, if for every distinct pair of vertices $x,y\in G$
with $x\neq y$ there is an $s\in S$ such that $s (x) \neq s (y)$;
\item[(f)] {\em full}, if for every nonadjacent pair of vertices $x,y\in G$
there is an $s \in S$ such that $s (x) = s (y) = 1$.
\end{itemize}
A full set of
two-valued states is separating,
and a separating set of
two-valued states is unital.
As will be detailed later TIFS/10-gadgets have a nonfull set so two-valued states in the sense of~(f).
Nonseparability in the sense of~(s) indicates nonclassicality.
And nonunitality in the sense of~(u) discredits the classical predictions of quantum clouds
even to a greater degree, probably only challenged by a complete absence of two-valued states.

\section{Chromatic separability}

As already discussed by Kochen and Specker~\cite[Theorem~0]{kochen1}
nonseparability of (at least one) pair of nonadjacent vertices
with respect to the set of two-valued states
(interpretable as classical truth assignments) of a graph
is arguably the most important signature of nonclassicality.
It may be true even if there is an ``abundance'' of two-valued states.
Nonseparability can also be expressed in terms of graph coloring.

 A proper (vertex)
coloring~\cite[Appendix]{Godsil-Newman-2008} of a graph is a function $c$ from the vertex set to a finite
set of ``colors'' (the positive integers will do) such that,
whenever $x$ and $y$ are adjacent vertices,
$c(x) \neq c(y)$.
The chromatic number of a graph  is the least positive integer $t$ such
that the graph has a coloring with $t$ colors.

A two-valued state on a (hyper)graph composed/pasted from contexts/cliques,
all having an identical
number of vertices/clique numbers $d$ can be obtained by ``projecting/reducing'' colors
if the chromatic number of that graph equals $d$.
In this case, in order to obtain a two-valued state,
take any proper (vertex) coloring $c$ with $d$ colors
and  map $d-1$ colors into (the ''new color'') $0$ and one color into (the ''new color'') $1$.
Just as the graph coloring $c$ itself such mappings need not to be unique.

Note that the chromatic number of a complete graph must be equal to the clique number
because type~(II)~{\&}~(III) two-valued states require that every context/clique
must have exactly one vertex with value assignment~$1$ (and~$0$ for all the other vertices).
(One might conjecture that the set of two-valued states
induces graph colorings, in much the same way as it induces a partition logic~\cite{svozil-2001-eua}.)
At the same time, the clique number renders a bound from below on the chromatic number.
Thus if the chromatic number of a graph exceeds the clique number
no such two-valued states exist -- the ``phenomenological consequence'' is that, for at least one context/clique,
the color projection/reduction is constant -- namely $0$ -- on all vertices of that context/clique.

A coloring is {\em chromatically separating} two nonadjacent distinct vertices $x$ and $y$ in the vertex set of a graph
if there exists a proper vertex coloring such that $c(x) \neq c(y)$.
A set of colorings of a particular graph is said to be {\em separating}
if, for any pair of distinct vertices it contains at least one coloring which separates those vertices.
The {\em separable chromatic number} of a graph is the least positive integer $t$ such
that the graph has a separating set of colorings with at most $t$ colors.

If the separable chromatic number is higher than the chromatic number of a given graph,
then there exist nonadjacent vertices which cannot be ``resolved''
by any proper graph coloring, or, for that matter, by any derived
projected/reduced two-valued state.
As a consequence, the graph has no set-theoretic realization as a partition logic,
although its chromatic number is the clique number,
and there still may exist an abundance of proper colorings and two-valued states~\cite[$\Gamma_3$, p.~70]{kochen1}.

It would be interesting to translate (non)unitality and (non)fullness into (sets of) graph coloring.
For reasons of brevity, we shall not discuss this here.

\section{Formation of gadgets as useful subgraphs for the construction of clouds}

The commonly used method seeks cloud configurations with ``exotic'' classical interpretations.
Again, exoticism may express itself in various forms or types.
One way is in terms of violations on bounds on classical {\em conditions of possible experience}~\cite[p.~229]{Boole-62},
such as Bell-type inequalities derivable from taking all [type~(II),
and type~(I) for inequalities using only the assumption of noncontextuality~\cite{cabello:210401}] value assignments,
forming a correlation polytope by encoding those states into vertices, and solving the hull problem
thereof~\cite{froissart-81,cirelson,pitowsky-86,pitowsky,pitowsky-89a,Pit-91,Pit-94,2000-poly}.
Another, stronger~\cite{svozil-2011-enough} form of nonclassicality is the nonexistence
of any such classical interpretation in terms of a type~(II) valued
assignments~\cite{Gleason,specker-60,Zierler1975,kochen1,pitowsky:218};
or at least their nonseparability~\cite[Theorem~0, p.~67]{kochen1}.

The explicit construction of such exotic classical interpretations
often proceeds by the (successive) application of exploitable subconfigurations of contexts
-- in graph theoretical terms {\em gadgets}~\cite{tutte_1954,SZABO2009436,Ramanathan-18} defined as ``useful subgraphs''.
Thereby, gadgets are formed from constituent lower order gadgets of ever-increasing size and functional performance
(see also~\cite[Chapter~12]{svozil-2016-pu-book}):
\begin{enumerate}

\item 0th order gadget:  a single context (aka clique/block/Boolean (sub)algebra/maximal observable/orthonormal basis).
This can be perceived as the most elementary form of a
true-implies-false (TIFS/10)~\cite{2018-minimalYIYS}/01-(maybe better 10)-gadget~\cite{svozil-2006-omni,Ramanathan-18}
configuration, because a truth/ value $1$ assignment of one of the vertices implies falsity/ value 0 assignments of all the others;

\item 1st order ``firefly'' gadget: two contexts connected in a single intertwining vertex;

\item 2nd order gadget:  two 1st order  firefly  gadgets connected in a single intertwining vertex;

\item 3rd order house/pentagon/pentagram gadget:  one firefly and one 2nd order gadget connected in two intertwining vertices to form a cyclic orthogonality hypergraph;

\item 4rth order 10-gadget:  e.g., a Specker bug~\cite{2018-minimalYIYS} consisting of two pentagon gadgets connected by an entire context;
as well as extensions thereof to arbitrary angles for the terminal (``extreme'') points~\cite{2015-AnalyticKS,Ramanathan-18};

\item 5th order  true-implies-true (TITS)~\cite{2018-minimalYIYS}/11-gadget~\cite{svozil-2006-omni}:  e.g.,  Kochen and Specker's $\Gamma_1$~\cite{kochen1}, consisting of one 10-gadget and one firefly gadget,
connected at the respective terminal points (cf.~Fig.~\ref{2018-c-fcloud10-11});

%
%
\end{enumerate}
That is, gadgets are subconfigurations of clouds. And clouds can be interpreted as gadgets for the composition of bigger clouds.

\begin{figure}
\begin{center}
\begin{tikzpicture}  [scale=0.25]

\tikzstyle{every path}=[line width=2pt]

\newdimen\ms
\ms=0.1cm

\tikzstyle{s1}=[color=red,rectangle,inner sep=3.5]
\tikzstyle{c3}=[circle,inner sep={\ms/8},minimum size=5*\ms]
\tikzstyle{c2}=[circle,inner sep={\ms/8},minimum size=3*\ms]
\tikzstyle{c1}=[circle,inner sep={\ms/8},minimum size=2*\ms]

\newdimen\R
\R=14.5cm     



\path
  ({180 + 0 * 360 /2}:\R      ) coordinate(1)
  ({180 + 1 * 360 /2}:\R   ) coordinate(2)
;



\draw [color=orange] (1) -- (2);

\node[cloud, cloud puffs=15.7, cloud ignores aspect, minimum width=4.5cm, minimum height=3cm, align=center, draw] (cloud) [color=black,fill=red!5]  at (0cm, 0cm) {\Large{\color{black}\vspace{1cm}  intertwined contexts \vspace{1cm} }};

%
%

\draw (1) coordinate[c3,fill=black];   %
\draw (1) coordinate[c2,fill=black!20,label=180:\colorbox{white}{\Large ${\bf a}$}];  %

\draw (2) coordinate[c3,fill=black];   %
\draw (2) coordinate[c2,fill=black!20,label=right:\colorbox{white}{\Large  ${\bf b}$}];  %
\end{tikzpicture}
\end{center}
\caption{
\label{2018-c-fcloud}
A collection of possible connections of counterfactuals organised in intertwining contexts and joining ${\bf a}$ and ${\bf b}$, depicted as a cloud
$C({\bf a},{\bf b})$.}
\end{figure}
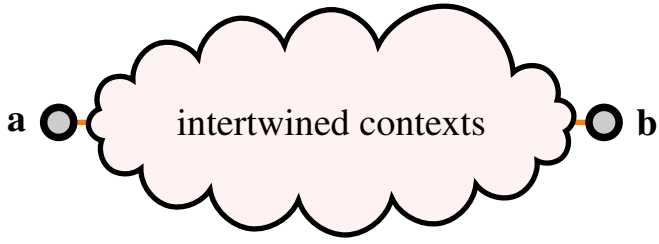

For the sake of arguing for an idealistic~\cite{berkeley,stace,Goldschmidt2017-idealism} and against a realistic usage of quantum clouds,
configurations of intertwined contexts with two fixed propositions
as ``start'' and ``end'' points ${\bf a}$ and ${\bf b}$ will be studied;
as well as methods for constructing such configurations with particular {\em relational} properties.
Whenever there is no preferred, less so unique, path connecting  ${\bf a}$ and ${\bf b}$, all such connections should be treated on an equal basis.
We shall call any such collection of counterfactual connections ``{\em clouds} connecting ${\bf a}$ and ${\bf b}$'',
denoted by $C({\bf a},{\bf b})$, and depict it with a cloud shape symbol,
as drawn in Figure~\ref{2018-c-fcloud}. (This can in principle be generalized to more than two terminal points.)

Thereby, as the endpoints ${\bf a}$ and ${\bf b}$ remain fixed, one can ask what kind of (classical) {\em relational information} can
be inferred from such two-point quantum clouds.
As it turns out, for fixed  ${\bf a}$ and ${\bf b}$ quantum clouds can be found which realize a wide variety
of conceivable relational properties between ${\bf a}$ and ${\bf b}$.
Table~\ref{2018-c-table1-conc-rel-info} enumerates these relations.
\begingroup
\squeezetable
\begin{table*}
 \begin{ruledtabular}
\begin{tabular}{lll}
\\
if ${\bf a}$ is true   & anectodal, historic    & reference to utility \\
classical value assignments  & quantum realisation & or relational properties \\
\\
\hline
\\
imply ${\bf b}$ is independent (arbitrary) & firefly logic $L_{12}$~\cite[pp.~21,~22]{cohen}&  \\
imply ${\bf b}$ false (TIFS/10)  & Specker bug logic~\cite[Fig.~1, p.~182]{kochen2} &   \cite[p.~588-589]{stairs83}, \cite{Yu-2012}, \cite{2018-minimalYIYS}\\
imply ${\bf b}$ true (TITS)  & extended Specker bug logic &  \cite[$\Gamma_1$, p.~68]{kochen1}, \\
&&\cite[Sects.~II,III, Fig.~1]{clifton-93}, \\
&&\cite[Fig.~C.l. p.~67]{Belinfante-73}, \\
&&\cite[p.~394]{Pitowsky-1982-subs}, \cite{Hardy-92,Hardy-93,hardy-97}, \\
&&\cite{Cabello-1995-ppks,cabello-96,cabello-97-nhvp,Badziag-2011,Cabello-2013-HP,Cabello-2013-Hardylike}, \cite{2018-minimalYIYS}\\
iff ${\bf b}$ true  (nonseparability) & combo of intertwined Specker bugs &  \cite[$\Gamma_3$, p.~70]{kochen1}\\
imply value indefiniteness of ${\bf b}$  & depending on Type (II), (III) assignments &  \cite{pitowsky:218}, \cite{2015-AnalyticKS}\\
\\
\end{tabular}
\end{ruledtabular}
\caption{Some (incomplete) history of the relational properties realizable by two-point quantum clouds.
\label{2018-c-table1-conc-rel-info}}
\end{table*}
\endgroup

\section{Quantum clouds enforcing particular features when interpreted classically}

For quantum mechanics, ${\bf a}$ and ${\bf b}$ can be
formalized by the two one dimensional projection operators
$\textsf{\textbf{E}}_{\bf a} =\vert {\bf a} \rangle \langle {\bf a}\vert$ and
$\textsf{\textbf{E}}_{\bf b} =\vert {\bf b} \rangle \langle {\bf b}\vert$, respectively.
For the sake of demonstration we shall study configurations in which
$\vert {\bf a} \rangle = \begin{pmatrix}1,0,0\end{pmatrix}$
and
$\vert {\bf b} \rangle = \begin{pmatrix}   \frac{1}{\sqrt{2}},\frac{1}{2},\frac{1}{2} \end{pmatrix}$,
that is, the quantum prediction yields a probability $\vert \langle {\bf b} \vert {\bf a} \rangle \vert^2 = \frac{1}{2}$
to find
the quantum in a state $\vert {\bf b} \rangle$ if it has been prepared in a state $\vert {\bf a}\rangle$.
This configuration can be extended to endpoints with (noncollinear and nonorthogonal) arbitrary relative location
by the techniques introduced in Refs.~\cite{2015-AnalyticKS,Ramanathan-18}.

\begin{enumerate}
\item[(a)]
A quantum cloud configuration for which classical value assignments allow
${\bf b}$ to be either true or false if  ${\bf a}$ is true
is the firefly configuration~\cite[pp.~21,~22]{cohen}, depicted in Fig.~\ref{2018-c-f-firefly},
with five classical value assignments of type (II)~\cite{dvur-pul-svo}.
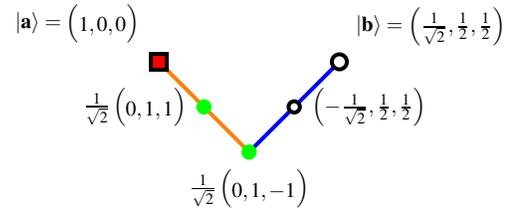
\begin{figure}
\begin{center}
\begin{tikzpicture}  [scale=0.20]

\newdimen\ms
\ms=0.05cm

\tikzstyle{every path}=[line width=1.5pt]

\tikzstyle{s1}=[color=red,rectangle,inner sep=3]
\tikzstyle{c4}=[circle,inner sep={\ms/8},minimum size=4*\ms]
\tikzstyle{c3}=[circle,inner sep={\ms/8},minimum size=3*\ms]
\tikzstyle{c2}=[circle,inner sep={\ms/8},minimum size=2*\ms]
\tikzstyle{c1}=[circle,inner sep={\ms/8},minimum size=1*\ms]

\newdimen\R
\R=6cm     



\path
  (0,6 ) coordinate(1)
  (3,3    ) coordinate(2)
  (6,0 ) coordinate(3)
  (9,3) coordinate(4)
  (12,6  ) coordinate(5)
;


\draw [color=orange] (1) -- (2) -- (3);
\draw [color=blue] (3) -- (4) -- (5);

%
%
\draw (1) coordinate[s1,draw=black,fill=red,label={above left: \footnotesize $\vert {\bf a} \rangle = \begin{pmatrix}1,0,0\end{pmatrix}$}];   %
\draw (2) coordinate[c3,draw=green,fill=green,label={left: \footnotesize $\frac{1}{\sqrt{2}}\begin{pmatrix}0,1,1\end{pmatrix}$}];    %
\draw (3) coordinate[c3,draw=green,fill=green,label={below: \footnotesize $\frac{1}{\sqrt{2}}\begin{pmatrix}0,1,-1\end{pmatrix}$}]; %
%
\draw (4) coordinate[c3,draw=black,fill=white,label={right: \footnotesize $\begin{pmatrix} -\frac{1}{\sqrt{2}},\frac{1}{2},\frac{1}{2}\end{pmatrix}$}];  %
\draw (5) coordinate[c4,draw=black,fill=white,label={above right: \footnotesize $\vert {\bf b} \rangle = \begin{pmatrix}   \frac{1}{\sqrt{2}},\frac{1}{2},\frac{1}{2} \end{pmatrix}$}];  %

\end{tikzpicture}
\end{center}
\caption{
\label{2018-c-f-firefly}
Orthogonality hypergraph of a cloud consisting of a
firefly logic $L_{12}$ connecting ${\bf a}$ and ${\bf b}$, such that, for type (II) value assignments, ${\bf a}$~true-implies-${\bf b}$~whatever (quantum 50:50).
Truth is encoded by a filled red square, classical falsity by a filled green circle, and arbitrary truth values by circles.
[Type (III) value assignments are partial and thus undefined.]
$L_{12}$ consists of 5~vertices in just 2~interetwined blocks allowing
a separating set of 5~two-valued states and therefore is set representable by partition logics.}
\end{figure}

\item[(b)]
Already Kochen and Specker utilized quantum clouds enforcing classical ${\bf a}$~true-implies-${\bf b}$~false  predictions
and their compositions in the construction of a configuration that does not allow a uniform truth assignment [of type (II)].
Stairs~\cite[p.~588-589]{stairs83} has pointed out that the Specker bug~\cite[Fig.~1, p.~182]{kochen2}
is a quantum cloud configuration which classically enforces ${\bf a}$~true-implies-${\bf b}$~false:
if a quantum system is prepared in such a way that ${\bf a}$ is true
-- that is, if it is in the state $\textsf{\textbf{E}}_{\bf a}$ -- and measured along $\textsf{\textbf{E}}_{\bf b}$, and
$\vert {\bf a} \rangle$ and
$\vert {\bf b} \rangle$
are not orthogonal or collinear, then any observation of ${\bf b}$ given ${\bf a}$ amounts to a probabilistic proof of nonclassicality:
because although quantum probabilities do not vanish, classical value assignments predict that ${\bf b}$ never occurs.
Minimal quantum cloud configurations for classical ${\bf a}$~true-implies-${\bf b}$~false,
as well as  ${\bf a}$~true-implies-${\bf b}$~true value assignments [of type (II)] can be found in~\cite{2018-minimalYIYS}.

As Cabello has pointed out~\cite{cabello-1994,Cabello-1996-diss}, the original Specker bug configuration cannot go beyond
the quantum prediction probability threshold $\vert \langle {\bf b} \vert {\bf a} \rangle \vert^2 = 3^{-2}$
because the angle between ${\bf a}$ and ${\bf b}$ cannot be smaller than
$\arccos {\frac{1}{3}} \approx 1.23096$ radians ($71.5^\circ$).
A configuration~\cite[Fig.~5(a)]{svozil-2018-whycontexts} allowing type (III) TIFS truth assignments
with ``maximally unbiased''
quantum prediction probability $\vert \langle {\bf b} \vert {\bf a} \rangle \vert^2 = \frac{1}{2}$
is a sublogic of a quantum logic
whose realization is enumerated in Ref.~\cite[Table.~1, p.~102201-7]{2015-AnalyticKS}.
It is depicted in Fig.~\ref{2018-c-f-TIFS}.
A proof of Theorem~2 in Ref.~\cite{Ramanathan-18} contains an explicit parametrization of a single
TIFS/10~cloud allowing the full range of angles $0 < \angle {\bf a},{\bf b} < \pi$.
\begin{figure}
\newif\iflabel \labelfalse
\begin{center}
\begin{tikzpicture}  [scale=0.25, rotate=117]

        \tikzstyle{every path}=[line width=1.5pt]
        \tikzstyle{c1}=[color=green,circle,inner sep=2.0]
        \tikzstyle{s1}=[color=red,rectangle,inner sep=2.5]
        \tikzstyle{l1}=[draw=none,circle,minimum size=4]


\draw [color=orange]   (4,0) coordinate[c1,draw=black,fill,label=0:{\color{black}\footnotesize $\vert {\bf b} \rangle$}] (b) -- (13,0)     coordinate[c1,fill,label=270:{\iflabel \tiny $P_2$\fi}] (2) -- (22,0)  coordinate[s1,fill,label=315:{\iflabel \tiny $P_3$\fi}] (3);
\draw [color=blue,   ] (3) -- (26,12)  coordinate[c1,fill,pos=0.8,label=0:{\iflabel \tiny $P_{21}$\fi}] (21) coordinate[c1,fill,label=0:{\iflabel \tiny $P_{23}$\fi}] (23);
\draw [color=white] (23) -- (22,18.5) coordinate[c1,fill,pos=0.4,color=white,label=0:{\iflabel \tiny $P_{29}$\fi}] (29) coordinate[c1,fill,label=45:{\iflabel \tiny $P_5$\fi}] (5);
\draw [color=magenta,] (5)-- (13,18.5)coordinate[s1,draw=black,fill,label=180:{\color{black}\footnotesize $\vert {\bf a} \rangle$}] (a) -- (4,18.5)  coordinate[c1,fill,label=135:{\iflabel \tiny $P_4$\fi}] (4);
\draw [color=CadetBlue, ] (4) -- (0,12)   coordinate[c1,fill,pos=0.6,label=180:{\iflabel \tiny $P_{10}$\fi}] (10) coordinate[s1,fill,label=180:{\iflabel \tiny $P_7$\fi}] (7);
\draw [color=brown,  ](7) -- (b)       coordinate[c1,fill,pos=0.2,label=180:{\iflabel \tiny $P_6$\fi}] (6);

        \draw [color=gray] (a) -- (2) coordinate[c1,fill,pos=0.5,label=315:{\iflabel \tiny $P_1$\fi}] (1);

        \draw [color=violet] (5) -- (22,6) coordinate[s1,fill,pos=0.4,label=0:{\iflabel \tiny $P_{11}$\fi}] (11) coordinate[c1,fill,label=0:{\iflabel \tiny $P_9$\fi}] (9);

\draw [color=Apricot] (9) -- (b) coordinate[s1,fill,pos=0.3,label=280:{\iflabel \tiny $P_8$\fi}] (8);

\draw [color=TealBlue] (4) -- (4,6) coordinate[s1,fill,pos=0.4,label=180:{\iflabel \tiny $P_{28}$\fi}] (28) coordinate[c1,fill,label=180:{\iflabel \tiny $P_{22}$\fi}] (22);
\draw [color=YellowGreen] (22) -- (3) coordinate[c1,fill,pos=0.2,label=260:{\iflabel \tiny $P_{19}$\fi}] (19);

        \coordinate (25) at ([xshift=-4cm]1);
        \coordinate (27) at ([xshift=4cm]1);

\draw [color=MidnightBlue]  (22) -- (25) coordinate[c1,fill,pos=0.5,label=115:{\iflabel \tiny $P_{24}$\fi}] (24) coordinate[s1,fill,label=270:{\iflabel \tiny $P_{25}$\fi}] (25);
\draw [color=Mulberry] (25) -- (9) coordinate[c1,fill,pos=0.8,label=90:{\iflabel \tiny $P_{35}$\fi}] (35);

\draw [color=BrickRed]  (7) -- (27) coordinate[c1,fill,pos=0.5,label=90:{\iflabel \tiny $P_{34}$\fi}] (34) coordinate[c1,fill,label=90:{\iflabel \tiny $P_{27}$\fi}] (27);
\draw [color=Emerald] (27) -- (23) coordinate[s1,fill,pos=0.25,label=270:{\iflabel \tiny $P_{26}$\fi}] (26);

\draw [color=BlueGreen]  (10) -- (15.5,17.5) coordinate[c1,fill,pos=0.5,label=90:{\iflabel \tiny $P_{12}$\fi}] (12) coordinate[s1,fill,label=15:{\iflabel \tiny $P_{13}$\fi}] (13);

\draw [color=RawSienna]  (28) -- (10.5,15) coordinate[c1,fill,pos=0.5,label=90:{\iflabel \tiny $P_{30}$\fi}] (30) coordinate[c1,fill,label=90:{\iflabel \tiny $P_{15}$\fi}] (15);
\draw [color=SpringGreen] (15) -- (11) coordinate[c1,fill,pos=0.6,label=90:{\iflabel \tiny $P_{14}$\fi}] (14);

\draw [color=Salmon]  (15) -- (1) coordinate[s1,fill,pos=0.2,label=15:{\iflabel \tiny $P_{17}$\fi}] (17);
\draw [color=Fuchsia] (1)-- (13) coordinate[c1,fill,draw=black,pos=0.3,label=90:{\color{black}\footnotesize  $\vert {16} \rangle$}] (16);

\draw [color=CornflowerBlue]  (19) -- (16) coordinate[s1,fill,pos=0.3,label=180:{\iflabel \tiny $P_{18}$\fi}] (18);
\draw [color=pink] (16) -- (8) coordinate[c1,fill,pos=0.7,label=180:{\iflabel \tiny $P_{32}$\fi}] (32);

\draw [color=PineGreen]  (6) -- (17) coordinate[c1,fill,pos=0.7,label=90:{\iflabel \tiny $P_{33}$\fi}] (33);
\draw [color=DarkOrchid] (17) -- (21) coordinate[c1,fill,pos=0.4,label=90:{\iflabel \tiny $P_{20}$\fi}] (20);

\draw [color=black] (25)  -- (1) -- (27);


\end{tikzpicture}
\end{center}
\caption{\label{2018-c-f-TIFS}
Orthogonality hypergraph of a nonfull/TIFS/10~cloud even for type (III) value assignments.
A faithful orthogonal realization is enumerated in Ref.~\cite[Table.~1, p.~102201-7]{2015-AnalyticKS}.
It consists of 38~vertices in 24~interetwined blocks,
endowed with a nonseparating set of 13~two-valued states and therefore is not set representable by partition logics.
The state depicted is the only one allowing ${\bf a}$ to be $1$.
Moreover, this cloud has no unital set of two-valued states as for all of them the vertex
represented by the vector
$
\vert {16} \rangle
=
\frac{1}{\sqrt{10}}
\left(
2\sqrt{2},1,-1
\right)
$ and
drawn as a solid black circle (and the associated
observable)  needs to be zero at all classical instantiations.
}
\end{figure}
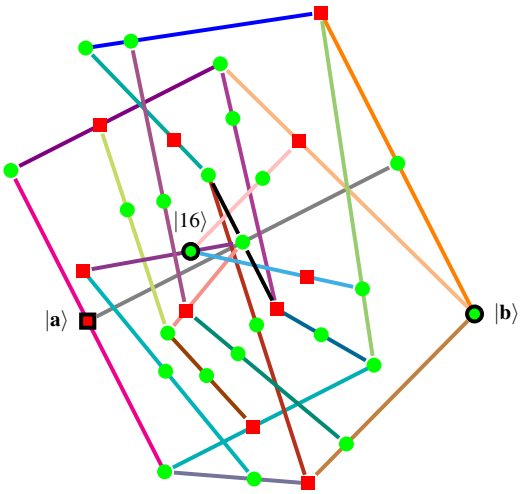


\item[(c)]
Clifton (note added in proof to Stairs~\cite[p.~588-589]{stairs83})
presented a ${\bf a}$~true-implies-${\bf b}$~true (TITS) cloud~\cite[Sects.~II,III, Fig.~1]{clifton-93,Johansen-1994,Vermaas-1994}
inspired by Bell~\cite[Fig.~C.l. p.~67]{Belinfante-73} (cf. also Pitowsky~\cite[p.~394]{Pitowsky-1982-subs}),
as well as by the Specker bug logic~\cite[Sects.~IV, Fig.~2]{clifton-93}.
Hardy~\cite{Hardy-92,Hardy-93,hardy-97}
as well as Cabello, among others~\cite{cabello-1994,Cabello-1996-diss,cabello-96,cabello-97-nhvp,Badziag-2011,Cabello-2013-HP,Cabello-2013-Hardylike,Ramanathan-18}
utilized similar scenarios for the demonstration of nonclassicality~\cite[Chapter~14]{svozil-pac}.
Fig.~\ref{2018-c-f-TITS} depicts a 11-gadget~\cite[Fig.~5(b)]{svozil-2018-whycontexts} with identical endpoints as the 10-gadget
discussed earlier and depicted in Fig.~\ref{2018-c-f-TIFS}.
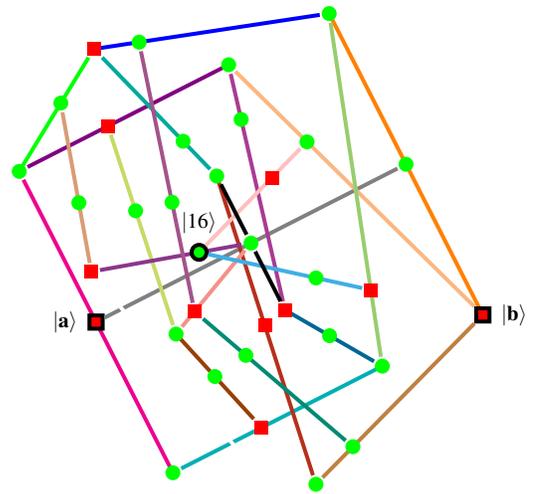
\begin{figure}
\newif\iflabel
\labelfalse
\begin{center}
\begin{tikzpicture}  [scale=0.25, rotate=117]

        \tikzstyle{every path}=[line width=1.5pt]
        \tikzstyle{c1}=[color=green,circle,inner sep=2.0]
        \tikzstyle{s1}=[color=red,rectangle,inner sep=2.5]
        \tikzstyle{l1}=[draw=none,circle,minimum size=4]


\draw [color=orange] (4,0) coordinate[s1,draw=black,fill,label=0:{\color{black}\footnotesize $\vert {\bf b}\rangle$}] (b) -- (13,0)  coordinate[c1,fill,label=270:{\iflabel \tiny $P_2$\fi}] (2) -- (22,0) coordinate[c1,fill,label=315:{\iflabel \tiny $P_3$\fi}] (3);
\draw [color=blue, ] (3) -- (26,12) coordinate[c1,fill,pos=0.8,label=0:{\iflabel \tiny $P_{21}$\fi}] (21) coordinate[s1,fill,label=0:{\iflabel \tiny $P_{23}$\fi}] (23);
\draw [color=green] (23) -- (22,18.5) coordinate[c1,fill,pos=0.4,label=0:{\iflabel \tiny $P_{29}$\fi}] (29) coordinate[c1,fill,label=45:{\iflabel \tiny $P_5$\fi}] (5);
\draw [color=magenta,] (5)-- (13,18.5)coordinate[s1,draw=black,fill,label=180:{\color{black}\footnotesize $\vert {\bf a}\rangle$}] (a) -- (4,18.5) coordinate[c1,fill,label=135:{\iflabel \tiny $P_4$\fi}] (4);
\draw [color=white] (4) -- (0,12) coordinate[c1,color=white,fill,pos=0.6,label=180:{\iflabel \tiny $P_{10}$\fi}] (10) coordinate[c1,fill,label=180:{\iflabel \tiny $P_7$\fi}] (7);
\draw [color=brown, ] (7) -- (b)   coordinate[c1,fill,pos=0.2,label=180:{\iflabel \tiny $P_6$\fi}] (6);

   \draw [color=gray] (a) -- (2) coordinate[c1,fill,pos=0.5,label=315:{\iflabel \tiny $P_1$\fi}] (1);

   \draw [color=violet] (5) -- (22,6) coordinate[s1,fill,pos=0.4,label=0:{\iflabel \tiny $P_{11}$\fi}] (11) coordinate[c1,fill,label=0:{\iflabel \tiny $P_9$\fi}] (9);

\draw [color=Apricot] (9) -- (b) coordinate[c1,fill,pos=0.3,label=280:{\iflabel \tiny $P_8$\fi}] (8);

\draw [color=TealBlue] (4) -- (4,6) coordinate[s1,fill,pos=0.4,label=180:{\iflabel \tiny $P_{28}$\fi}] (28) coordinate[c1,fill,label=180:{\iflabel \tiny $P_{22}$\fi}] (22);
\draw [color=YellowGreen] (22) -- (3) coordinate[s1,fill,pos=0.2,label=260:{\iflabel \tiny $P_{19}$\fi}] (19);

   \coordinate (25) at ([xshift=-4cm]1);
   \coordinate (27) at ([xshift=4cm]1);

\draw [color=MidnightBlue] (22) -- (25) coordinate[c1,fill,pos=0.5,label=115:{\iflabel \tiny $P_{24}$\fi}] (24) coordinate[s1,fill,label=270:{\iflabel \tiny $P_{25}$\fi}] (25);
\draw [color=Mulberry] (25) -- (9) coordinate[c1,fill,pos=0.8,label=90:{\iflabel \tiny $P_{35}$\fi}] (35);

\draw [color=BrickRed] (7) -- (27) coordinate[s1,fill,pos=0.5,label=90:{\iflabel \tiny $P_{34}$\fi}] (34) coordinate[c1,fill,label=90:{\iflabel \tiny $P_{27}$\fi}] (27);
\draw [color=Emerald] (27) -- (23) coordinate[c1,fill,pos=0.25,label=270:{\iflabel \tiny $P_{26}$\fi}] (26);

\draw [color=white] (10) -- (15.5,17.5) coordinate[c1,color=white,fill,pos=0.5,label=90:{\iflabel \tiny $P_{12}$\fi}] (12) coordinate[s1,fill,label=15:{\iflabel \tiny $P_{13}$\fi}] (13);
\draw [color=Tan] (13) -- (29) coordinate[c1,fill,pos=0.4,label=90:{\iflabel \tiny $P_{31}$\fi}] (31);

\draw [color=RawSienna] (28) -- (10.5,15) coordinate[c1,fill,pos=0.5,label=90:{\iflabel \tiny $P_{30}$\fi}] (30) coordinate[c1,fill,label=90:{\iflabel \tiny $P_{15}$\fi}] (15);
\draw [color=SpringGreen] (15) -- (11) coordinate[c1,fill,pos=0.6,label=90:{\iflabel \tiny $P_{14}$\fi}] (14);

\draw [color=Salmon] (15) -- (1) coordinate[s1,fill,pos=0.2,label=15:{\iflabel \tiny $P_{17}$\fi}] (17);
\draw [color=Fuchsia] (1)-- (13) coordinate[c1,fill,draw=black,pos=0.3,label=90:{\color{black}\footnotesize  $\vert {16} \rangle$}] (16);

\draw [color=CornflowerBlue] (19) -- (16) coordinate[c1,fill,pos=0.3,label=180:{\iflabel \tiny $P_{18}$\fi}] (18);
\draw [color=pink] (16) -- (8) coordinate[s1,fill,pos=0.7,label=180:{\iflabel \tiny $P_{32}$\fi}] (32);

\draw [color=PineGreen] (6) -- (17) coordinate[c1,fill,pos=0.7,label=90:{\iflabel \tiny $P_{33}$\fi}] (33);
\draw [color=DarkOrchid] (17) -- (21) coordinate[c1,fill,pos=0.4,label=90:{\iflabel \tiny $P_{20}$\fi}] (20);

\draw [color=black] (25) -- (1) -- (27);

   \coordinate (ContextLabel) at ([shift=({-2cm,-3mm})]1);
   \draw (ContextLabel) coordinate[l1,label=90:{\iflabel \tiny $C_{26}$\fi}];

   \end{tikzpicture}
\end{center}
\caption{
\label{2018-c-f-TITS}
Orthogonality hypergraph of a TITS/11 cloud even for type (III) value assignments.
A faithful orthogonal realization is enumerated in Ref.~\cite[Table.~1, p.~102201-7]{2015-AnalyticKS}.
It consists of 38~vertices in 24~interetwined blocks,
endowed with a nonseparating set of 13~two-valued states and therefore is not set representable by partition logics.
The state depicted is the only one allowing ${\bf a}$ to be $1$.
Moreover, this cloud has no unital set of two-valued states as for all of them the vertex
represented by the vector $\vert {16} \rangle = \frac{1}{\sqrt{10}}\left( 2\sqrt{2},1,-1 \right)$ and
drawn as a solid black circle (and the associated
observable)  needs to be zero at all classical instantiations.
}
\end{figure}

\item[(d)]
Various parallel and serial compositions of 10- and 11-gadgets serve as a ``gadget toolbox'' to obtain clouds which,  if they are interpreted classically,
exhibit other interesting relational properties.
For instance, the parallel composition (pasting) of two quantum clouds of the 10-gadget type:
one 10-gadget classically demanding ${\bf a}$~true-implies-${\bf b}$~false
and the other 10-gadget  classically demanding ${\bf b}$~true-implies-${\bf a}$~false, results in a  quantum cloud which  has two observables ${\bf a}$ and ${\bf b}$
which are classically always ``opposite'': if one is true, the other one is false, and {\it vice versa}.

\item[(e)]
The parallel composition (pasting) of two quantum clouds of the TITS type, with one TITS, classically demanding ${\bf a}$~true-implies-${\bf b}$~true
and the other TITS  classically demanding ${\bf b}$~true-implies-${\bf a}$~true, results in a  quantum cloud which  has two observables ${\bf a}$ and ${\bf b}$
which are classically nonseparable, which is a sufficient criterion for nonclassicality~\cite[Theorem~0, p.~67]{kochen1}.
As pointed out by Portillo~\cite{Portillo-2018-pc} this is equivalent to ${\bf a}$ is true if and only if ${\bf b}$ is true (TIFFTS).
Fig.~\ref{2018-c-f-TIFFTS} depicts a historic example of such a construction.
The serial composition of suitable TITS
of the form ${\bf a}_1$~true-implies-${\bf a}_2 \cdots {\bf a}_{i-1}$~true-implies-${\bf a_i}$~true
eventually yields two or more vectors ${\bf a}_1$ and ${\bf a_i}$ which are mutually orthogonal; a technique employed by Kochen and Specker
for the construction of a quantum cloud admitting no type (II) truth assignment~\cite[$\Gamma_2$, p.~69]{kochen1}.

\begin{figure}
\begin{center}
\begin{tikzpicture}  [scale=0.8, rotate= 0]

\newdimen\E
\E=0.05 cm

\tikzstyle{every path}=[line width=1.5pt]

\newdimen\ms
\ms=0.15 cm

        \tikzstyle{c1}=[color=green,circle,inner sep=2.5]
        \tikzstyle{s1}=[color=red,rectangle,inner sep=2.5]

\path
 (1,0)           coordinate(1)
 (0,1)           coordinate(2)
 (2,1)           coordinate(3)
 ({1 cm-2*\E},2) coordinate(4)
 ({1 cm-1*\E},2) coordinate(5)
 ({1 cm+1*\E},2) coordinate(6)
 ({1 cm+2*\E},2) coordinate(7)
 ({1 cm-2*\E},3) coordinate(8)
 ({1 cm-1*\E},3) coordinate(9)
 ({1 cm+1*\E},3) coordinate(10)
 ({1 cm+2*\E},3) coordinate(11)
 ({1 cm-2*\E},4) coordinate(12)
 ({1 cm-1*\E},4) coordinate(13)
 ({1 cm+1*\E},4) coordinate(14)
 ({1 cm+2*\E},4) coordinate(15)
 (0.5,5)         coordinate(16)
 (1.5,5)         coordinate(17)
 (0,6)           coordinate(18)
 (2,6)           coordinate(19)
 (1,7)           coordinate(20)
 (2.25,8.5)        coordinate(21)
 (4,8.5)         coordinate(22)
 (2.45,0.25)        coordinate(23) 
 (1.5,8)        coordinate(24) 
 (1,2)        coordinate(25) 
 (1,3)        coordinate(26) 
 (1,4)        coordinate(27) 

 (3,0)+(1,8.5)           coordinate(101)
 (3,0)+(0,7.5)           coordinate(102)
 (3,0)+(2,7.5)           coordinate(103)
 (3,0)+({1 cm-2*\E},6.5) coordinate(104)
 (3,0)+({1 cm-1*\E},6.5) coordinate(105)
 (3,0)+({1 cm+1*\E},6.5) coordinate(106)
 (3,0)+({1 cm+2*\E},6.5) coordinate(107)
 (3,0)+({1 cm-2*\E},5.5) coordinate(108)
 (3,0)+({1 cm-1*\E},5.5) coordinate(109)
 (3,0)+({1 cm+1*\E},5.5) coordinate(110)
 (3,0)+({1 cm+2*\E},5.5) coordinate(111)
 (3,0)+({1 cm-2*\E},4.5) coordinate(112)
 (3,0)+({1 cm-1*\E},4.5) coordinate(113)
 (3,0)+({1 cm+1*\E},4.5) coordinate(114)
 (3,0)+({1 cm+2*\E},4.5) coordinate(115)
 (3,0)+(0.5,3.5)         coordinate(116)
 (3,0)+(1.5,3.5)         coordinate(117)
 (3,0)+(0,2.5)           coordinate(118)
 (3,0)+(2,2.5)           coordinate(119)
 (3,0)+(1,1.5)           coordinate(120)
 (3,0)+(-0.25,0)        coordinate(121)
 (1,0)             coordinate(122)
 (3,0)+(-0.45,8.25)        coordinate(123) 
 (3.5,0.5)        coordinate(124) 
 (4,6.5)        coordinate(125) 
 (4,5.5)        coordinate(126) 
 (4,4.5)        coordinate(127) 
;


\draw [color=green,rounded corners=3mm] (1) -- (2);
\draw [color=violet, rounded corners=3mm] (1) -- (3);
\draw [color=gray, rounded corners=3mm] (2) -- (4) -- (8) -- (12) -- (16) -- (18);
\draw [color=magenta, rounded corners=3mm] (3) -- (7) -- (11) -- (15) -- (17) -- (19);
\draw [color=blue, rounded corners=3mm] (18) -- (20);
\draw [color=pink, rounded corners=3mm] (19) -- (20);
\draw [color=red, rounded corners=3mm] (1) -- (6) -- (10) -- (14);
\draw [color=lime, rounded corners=3mm]  (5) -- (9) -- (13) -- (20) -- (24) -- (21) -- (22);
\draw [color=orange, rounded corners=10mm] (1) -- (23) -- (21);
\draw [color=cyan, rounded corners=3mm] (16) -- (17);

\draw [color=green!50,rounded corners=3mm] (101) -- (102);
\draw [color=violet!50, rounded corners=3mm] (101) -- (103);
\draw [color=gray!50, rounded corners=3mm] (102) -- (104) -- (108) -- (112) -- (116) -- (118);
\draw [color=magenta!50, rounded corners=3mm] (103) -- (107) -- (111) -- (115) -- (117) -- (119);
\draw [color=blue!50, rounded corners=3mm] (118) -- (120);
\draw [color=pink!50, rounded corners=3mm] (119) -- (120);
\draw [color=red!50, rounded corners=3mm] (101) -- (106) -- (110) -- (114);
\draw [color=lime!50, rounded corners=3mm]  (105) -- (109) -- (113) -- (120) -- (124) -- (121) -- (122);
\draw [color=orange!50, rounded corners=10mm] (101) -- (123) -- (121);
\draw [color=cyan!50, rounded corners=3mm] (116) -- (117);

 \draw (1) coordinate[s1,draw=black,fill,label=180:{\color{black}\footnotesize $\vert {\bf a}\rangle$}];
 \draw (2) coordinate[c1,fill];
 \draw (3) coordinate[c1,fill];
 \draw (25) coordinate[c1,draw=green,fill];
 \draw (26) coordinate[c1,draw=green,fill];
 \draw (27) coordinate[c1,draw=green,fill];
\draw (16) coordinate[c1,draw=black,fill=white];
\draw (17) coordinate[c1,draw=black,fill=white];
\draw (18) coordinate[c1,draw=black,fill=white];
\draw (19) coordinate[c1,draw=black,fill=white];
\draw (20) coordinate[c1,fill];
 \draw (21) coordinate[c1,fill,label=180:{\color{black}\footnotesize $\vert {\bf c}\rangle$}];
\draw (101) coordinate[s1,draw=black,fill,label=0:{\color{black}\footnotesize $\vert {\bf b}\rangle$}];
\draw (102) coordinate[c1,fill];
\draw (103) coordinate[c1,fill];
\draw (125) coordinate[c1,draw=green,fill];
\draw (126) coordinate[c1,draw=green,fill];
\draw (127) coordinate[c1,draw=green,fill];
 \draw (116) coordinate[c1,draw=black,fill=white];
 \draw (117) coordinate[c1,draw=black,fill=white];
 \draw (118) coordinate[c1,draw=black,fill=white];
 \draw (119) coordinate[c1,draw=black,fill=white];
 \draw (120) coordinate[c1,fill];
\draw (121) coordinate[c1,fill,label=0:{\color{black}\footnotesize $\vert {\bf c}'\rangle$}];

\end{tikzpicture}
\end{center}
\caption{
\label{2018-c-f-TIFFTS}
Orthogonality hypergraph of a TIFFTS cloud for type (II) value assignments,
based on a minimal 11-gadgets introduced in Ref.~\cite[Fig.~6]{2018-minimalYIYS} for dimensions greater than 2.
In three dimensions,
(i) the three orthogonal ``middle'' vertices intertwining four contexts vanish,
(ii) the two vertices $\vert {\bf c}\rangle$ and $\vert {\bf c}'\rangle$  coincide, and
(iii) the two edges connecting $\vert {\bf c}\rangle$ with $\vert {\bf a}\rangle$  and $\vert {\bf c}'\rangle$
with $\vert {\bf b}\rangle$ vanish,
rendering the original Specker bug combo introduced by Kochen and Specker~\cite[$\Gamma_3$, p.~70]{kochen1}.
Unlike the earlier configurations, this cloud does not allow 50:50 quantum probabilities.
Because of nonseparability of its set of two-valued states
and its separable chromatic number higher than the clique number
it does not allow a set representation by partition logics.
}
\end{figure}
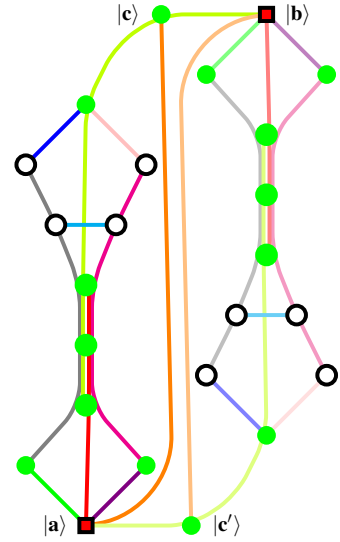

\item[(f)]
The parallel composition (pasting) of the two quantum clouds which respectively represent a 10-gadget and an 11-gadget
and identical endpoints   ${\bf a}$ and ${\bf b}$ yields a ${\bf a}$~true-implies-${\bf b}$~value indefinite cloud discussed in Ref.~\cite{2015-AnalyticKS}.

\end{enumerate}

\section{Some technical issues of gadget construction}

The concatenation of intertwining gadgets needs to allow a proper faithful orthogonal representation
of the resulting compound (hyper)graph while at the same time preserving the structure of these gadgets.
Thereby the faithful orthogonal representations of the constituent gadgets cannot always be transferred
easily to a faithful orthogonal representation of the resulting compound (hyper)graph.

Suppose, for the sake of a counterexample involving duplicity of vertices after concatenations of gadgets, one would attempt to construct a
$G
\left(
\begin{pmatrix}
1,0,0
\end{pmatrix}
,
\begin{pmatrix}
0,1,1
\end{pmatrix}
\right)
$
11~cloud (which would constitute a Kochen-Specker proof as the respective terminal points are orthogonal)
by concatenating two 11-gadgets
$
G
\left(
\begin{pmatrix}
1,0,0
\end{pmatrix}
,
\begin{pmatrix}
 \frac{1}{\sqrt{2}},\frac{1}{2},\frac{1}{2}
\end{pmatrix}
\right)
$
and $
G
\left(
\begin{pmatrix}
 \frac{1}{\sqrt{2}},\frac{1}{2},\frac{1}{2}
\end{pmatrix}
,
\begin{pmatrix}
0,1,1
\end{pmatrix}
\right)
$
of the type depicted in Fig.~\ref{2018-c-f-TITS}
by simply rotating all coordinates of the first gadget $\frac{\pi}{4}$ radians ($45^\circ$) about the axis formed by
${\bf b}-{\bf a}$.
Unfortunately, a straightforward calculation shows that these two 11-gadgets, with the faithful orthogonal realization taken from~\cite[Table~I, p.~102201-7]{2015-AnalyticKS},
do not only have the vertex $\begin{pmatrix}
 \frac{1}{\sqrt{2}},\frac{1}{2},\frac{1}{2}
\end{pmatrix}$
in common as per construction, but also the three additional vertices
$
\begin{pmatrix}
0,\frac{1}{\sqrt{2}},\pm \frac{1}{\sqrt{2}}
\end{pmatrix}
$
and
 $\begin{pmatrix}
 1,0,0
\end{pmatrix}
$.

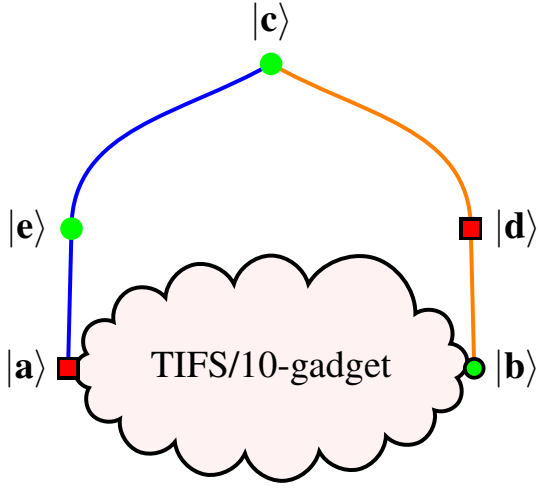
\begin{figure}
\begin{center}
\begin{tikzpicture}  [scale=0.20]

\newdimen\ms
\ms=0.1cm

        \tikzstyle{every path}=[line width=1.5pt]
        \tikzstyle{c3}=[fill=black,circle,inner sep={\ms/8},minimum size=3*\ms]
        \tikzstyle{c1}=[fill=green,circle,inner sep={\ms/8},minimum size=2*\ms]
        \tikzstyle{s3}=[fill=black,rectangle,inner sep={\ms/8},minimum size=3*\ms]
        \tikzstyle{s1}=[fill=red,rectangle,inner sep={\ms/8},minimum size=2*\ms]
        \tikzstyle{l1}=[draw=none,circle,minimum size=4]

\newdimen\R
\R=13.5cm     



\path
  ({180 + 0 * 360 /2}:\R      ) coordinate(1)
  ({180 + 1 * 360 /2}:\R   ) coordinate(2)
   ( {90}:{1.5 * \R} ) coordinate(3)
   ( {35}:{1.2 * \R} ) coordinate(4)
   ( {145}:{1.2 * \R} ) coordinate(5)
;





\draw [color=orange] (1) -- (2);

\draw [color=blue] (1) -- (5) to   [out=90,in=210] (3);
\draw [color=orange] (2) -- (4) to   [out=90,in=330] (3);

\node[cloud, cloud puffs=15.7, cloud ignores aspect, minimum width=5cm, minimum height=3cm, align=center, draw] (cloud) [color=black,fill=red!5]  at (0cm, 0cm) {\Large{\color{black}\vspace{1cm}  TIFS/10-gadget \vspace{1cm} }};

%
%

\draw (1) coordinate[s3,fill=black];   %
\draw (1) coordinate[s1,label=180:{\Large $\vert {\bf a}\rangle$}];  %

\draw (2) coordinate[c3];   %
\draw (2) coordinate[c1,label=0:{\Large  $\vert {\bf b}\rangle$}];  %

\draw (3) coordinate[c3,fill=green,label=90:{\Large  $\vert {\bf c}\rangle$}];
%
\draw (4) coordinate[s3,fill=black,label=0:{\Large  $\vert {\bf d}\rangle$}];
\draw (4) coordinate[s1,fill=red];

\draw (5) coordinate[c3,fill=green,label=180:{\Large  $\vert {\bf e}\rangle$}];
\end{tikzpicture}
\end{center}
\caption{
\label{2018-c-fcloud10-11}
Standard construction used by Kochen and Specker~\cite[Lemma~1, $\Gamma_1$, p.~68]{kochen1}
for obtaining a 11~cloud $C({\bf a},{\bf d})$ [or, because of symmetry, $C({\bf b},{\bf e})$]
from a nonfull/TIFS/10-gadget $C({\bf a},{\bf b})$,
involving two additional contexts $\left\{ {\bf b},{\bf d},{\bf c} \right\}$
and $\left\{ {\bf a},{\bf e} ,{\bf c}  \right\}$.}
\end{figure}

Also, gadgets may not be able to perform as desired.
For instance, a standard construction in three dimensions, already used by Kochen and Specker~\cite[Lemma~1, $\Gamma_1$, p.~68]{kochen1} for their construction of a
11-gadget $\Gamma_1$ from a Specker bug-type 10-gadget introduced earlier~\cite[Fig.~1, p.~182]{kochen2},
is to take the terminal points  ${\bf a}$ and ${\bf b}$ of some TIFS/10~cloud
and form the normal vector  ${\bf c} = {\bf a} \times {\bf b}$.
In a second step, the vector
\begin{equation}
\begin{split}
{\bf d} = {\bf b} \times {\bf c}
={\bf b} \times \left({\bf a} \times {\bf b}\right)      \\
  = {\bf b}^2  {\bf a} - \left({\bf a}\cdot {\bf b}\right)  {\bf b}
=  {\bf a} - \left({\bf a}\cdot {\bf b}\right)  {\bf b}
  =  {\bf a} - \cos \left( \angle {\bf a}, {\bf b}\right)  {\bf b}
\end{split}
\end{equation}
orthogonal to both ${\bf b}$ and ${\bf c}$ is formed.
If  ${\bf a}$ is true/1 then ${\bf b}$ (because of the 01-gadget) as well as ${\bf c}$ (because of orthogonality with ${\bf a}$) must be false/0.
Therefore ${\bf d}$ must be true, since it completes the context $\left\{ {\bf b},{\bf c},{\bf d}\right\}$.
The situation is depicted in Fig.~\ref{2018-c-fcloud10-11}.
If all goes well the new cloud $C({\bf a},{\bf d})$ is of the TITS/11~type.
This is not the case if one uses the TIFS/10-gadget depicted in Fig.~\ref{2018-c-f-TIFS},
as the vector
 ${\bf c} =
\begin{pmatrix}
0,
-\frac{1}{\sqrt{2}},
\frac{1}{\sqrt{2}}
\end{pmatrix}$
and the new terminal vector ${\bf d} =\begin{pmatrix}
 \frac{1}{\sqrt{2}},-\frac{1}{2},-\frac{1}{2}
\end{pmatrix}$
also appear in the original TIFS/10-gadget.

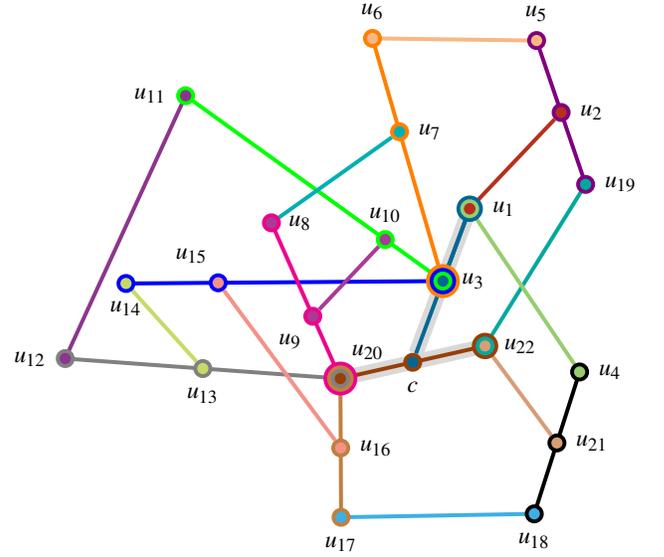
\begin{figure}
\begin{center}
\begin{tikzpicture}  [scale=0.5,rotate=0]

\tikzstyle{every path}=[line width=1.5pt]

\tikzstyle{c4}=[circle,inner sep=4pt,minimum size=7pt]
\tikzstyle{c3}=[circle,inner sep=3pt,minimum size=5pt]
\tikzstyle{c2}=[circle,inner sep=2pt,minimum size=3pt]
\tikzstyle{c1}=[circle,inner sep=1.5pt,minimum size=2pt]
\tikzstyle{c0}=[circle,inner sep=0.5pt,minimum size=1pt]


\path
( {4 *2.68838},{4 * 2.05247})   coordinate(1)
( {4 *3.41896},{4 *    0.966553})   coordinate(4)
( {4 *3.13324},{4 * 3.17128})   coordinate(5)
( {4 *2.04217},{4 * 3.18563})   coordinate(6)
( {4 *1.37051},{4 * 1.95877})   coordinate(8)
( {4 *0.8},{4 * 2.80416})   coordinate(11)
( {4 *0.00},{4 * 1.05577})   coordinate(12)
( {4 *0.40372},{4 * 1.55514})   coordinate(14)
( {4 *1.26923},{4 *    0.189747})   coordinate(16)
( {4 *1.83338},{4 * 0.})   coordinate(17)
( {4 *3.11765},{4 * 0.0226601})   coordinate(18)
( {4 *3.45895},{4 *    2.21445})   coordinate(19)
( {4 *1.82887},{4 * 0.922707})   coordinate(20)
( {4 *3.58461},{4 * 0.431072})   coordinate(21)
( {4 *2.79057},{4 *    1.13936})   coordinate(22)
;

\draw [line width=2mm,color=gray!30] (20) --    (22);
\draw [color=RawSienna]           (20) --    (22) coordinate[c2,fill=RawSienna,draw=RawSienna,pos=0.5,label=270:{\color{black}$c$}]    (31);
\draw [line width=2mm,color=gray!30] (1) --    (31);
\draw [color=MidnightBlue]           (1) --    (31) coordinate[c2,fill=MidnightBlue,draw=MidnightBlue,pos=0.5,label=0:{\color{black}$u_3$}]    (3);
\draw [color=orange]              (3) --  (6) coordinate[c2,fill=orange,draw=orange,pos=0.6,label=0:{\color{black}$u_{7}$}]    (7);
\draw [color=blue]              (3) -- (14) coordinate[c2,fill=blue,draw=blue,pos=0.7,label=95:{\color{black}$u_{15}$}]   (15);
\draw [color=green]               (3) --  (11)  coordinate[c2,fill=green,draw=green,pos=0.2]  (10);
\draw [color=magenta]            (8)  --  (20)  coordinate[c2,fill=magenta,draw=magenta,pos=0.6]   (9);
\draw [color=black]               (4) -- (18) coordinate[c2,fill=black,draw=black,pos=0.5,label=0:{\color{black}$u_{21}$}]   (21);
\draw [color=brown]             (17) --  (20) coordinate[c2,fill=brown,draw=brown,pos=0.5,label=0:{\color{black}$u_{16}$}] (16);
\draw [color=gray]                (12) --  (20)   coordinate[c2,fill=gray,draw=gray,pos=0.5,label=270:{\color{black}$u_{13}$}]  (13);
\draw [color=violet]              (5) --  (19) coordinate[c2,fill=violet,draw=violet,pos=0.5,label=0:{\color{black}$u_{2}$}]   (2);
\draw [color=Apricot]             (5) -- (6)  ;
\draw [color=TealBlue]            (7) --    (8);
\draw [color=MidnightBlue]        (3) --   (1);
\draw [color=Mulberry]            (10) --    (9);
\draw [color=BrickRed]            (2) --    (1);
\draw [color=Emerald]             (19) --    (22);
\draw [color=YellowGreen]          (1) --    (4);
\draw [color=Tan]                 (22) --    (21);
\draw [color=SpringGreen]         (14) --    (13);
\draw [color=Salmon]              (15) --    (16);
\draw [color=Fuchsia]             (11) --    (12);
\draw [color=CornflowerBlue]        (17) --    (18);

 \draw (10) coordinate[c1,fill=Mulberry,label=90:{\color{black}$u_{10}$}];
 \draw (13) coordinate[c1,fill=SpringGreen];
 \draw (15) coordinate[c1,fill=Salmon];
 \draw (9) coordinate[c1,fill=Mulberry,label=260:{\color{black}$u_{9}$}];
 \draw (7) coordinate[c1,fill=TealBlue];
 \draw (2) coordinate[c1,fill=BrickRed];
 \draw (21) coordinate[c1,fill=Tan];
 \draw (16) coordinate[c1,fill=Salmon];

 \draw (11) coordinate[c2,fill=green,draw=green,label=180:{\color{black}$u_{11}$}];
 \draw (11) coordinate[c1,fill=Fuchsia];

 \draw (12) coordinate[c2,fill=gray,draw=gray,label=180:{\color{black}$u_{12}$}];
 \draw (12) coordinate[c1,fill=Fuchsia];

 \draw (14) coordinate[c2,fill=blue,draw=blue];
 \draw (14) coordinate[c1,fill=SpringGreen,label=270:{\color{black}$u_{14}$}];

 \draw (3) coordinate[c4,fill=orange,draw=orange];
 \draw (3) coordinate[c3,fill=blue,draw=blue];
 \draw (3) coordinate[c2,fill=green,draw=green];
 \draw (3) coordinate[c1,fill=MidnightBlue];

 \draw (20) coordinate[c4,fill=magenta,draw=magenta];
 \draw (20) coordinate[c3,fill=brown,draw=brown];
 \draw (20) coordinate[c2,fill=gray,draw=gray,label=80:{\color{black}$u_{20}$}];
 \draw (20) coordinate[c1,fill=RawSienna];

 \draw (6) coordinate[c2,fill=orange,draw=orange,label=90:{\color{black}$u_{6}$}];
 \draw (6) coordinate[c1,fill=Apricot];

 \draw (5) coordinate[c2,fill=violet,draw=violet,label=90:{\color{black}$u_{5}$}];
 \draw (5) coordinate[c1,fill=Apricot];

 \draw (19) coordinate[c2,fill=violet,draw=violet,label=0:{\color{black}$u_{19}$}];
 \draw (19) coordinate[c1,fill=Emerald];

 \draw (4) coordinate[c2,fill=black,draw=black,label=0:{\color{black}$u_{4}$}];
 \draw (4) coordinate[c1,fill=YellowGreen];

 \draw (18) coordinate[c2,fill=black,draw=black,label=270:{\color{black}$u_{18}$}];
 \draw (18) coordinate[c1,fill=CornflowerBlue];

 \draw (17) coordinate[c2,fill=brown,draw=brown,label=270:{\color{black}$u_{17}$}];
 \draw (17) coordinate[c1,fill=CornflowerBlue];

 \draw (8) coordinate[c2,fill=magenta,draw=magenta];
 \draw (8) coordinate[c1,fill=Mulberry,label=0:{\color{black}$u_{8}$}];

 \draw (1) coordinate[c3,fill=MidnightBlue,draw=MidnightBlue,label=0:{\color{black}$u_{1}$}];
 \draw (1) coordinate[c2,fill=violet,draw=YellowGreen];
 \draw (1) coordinate[c1,fill=BrickRed];

 \draw (22) coordinate[c3,fill=RawSienna,draw=RawSienna];
 \draw (22) coordinate[c2,fill=Emerald,draw=Emerald,label=0:{\color{black}$u_{22}$}];
 \draw (22) coordinate[c1,fill=Tan];

 \draw (31) coordinate[c1,fill=MidnightBlue];

\end{tikzpicture}
\end{center}
\caption{\label{2019-c-HH10}
Orthogonality hypergraph from a proof of
Theorem~3 in Ref.~\cite{Ramanathan-18}.
The advantage of this nonfull/TIFS/10-gadget
is a straightforward parametric faithful orthogonal representation allowing angles
$0< \angle u_1,u_{22} \le \frac{\pi}{4}$ radians ($45^\circ$)
of, say, the terminal points $u_1$ and $u_{22}$.
The corresponding logic including the completed set of
34~vertices in
21~blocks is set representable by partition logics because the supported 89~two-valued
states are (color) separable.
It is not too difficult to prove (by contradiction) that, say,
if both $u_1$ as well as $u_{22}$
are assumed to be $1$, then
$u_2$,
$u_3$,
$u_4$,
as well as
$u_{19}$,
$u_{20}$
and $u_{21}$
should be $0$.
Therefore,
$u_5$ and
$u_{18}$
would need to be true.
As a result,
$u_{6}$ and
$u_{17}$ would need to be false.
Hence,
$u_{7}$ as well as
$u_{16}$ would be $1$,
rendering
$u_{8}$ and
$u_{15}$ to be $0$.
This would imply
$u_{9}$ as well as
$u_{14}$ to be $1$,
which in turn would demand
$u_{10}$ and
$u_{13}$ to be false.
Therefore,
$u_{11}$ and
$u_{12}$ would have to be $1$, which yields a complete contradiction even for type-III
value assignments.
It is also a TITS/11-gadget for the terminal points
$u_1-u_{20}$, constructed by the standard construction depicted in Fig.~\ref{2018-c-fcloud10-11}.
}
\end{figure}

For very similar reasons (degeneracy or division through zero)
the 10-gadget introduced in the proof of Theorem~3 in Ref.~\cite{Ramanathan-18}
and depicted in
Fig.~\ref{2019-c-HH10}
cannot be extended to an 11~cloud whose end terminals
are the ``maximal'' angle $\frac{\pi}{4}$ radians ($45^\circ$) apart.
For all other allowed angles an extension of this earlier construction
of a TIFS/10-gadget to a TITS/11-cloud
depicted in Fig.~\ref{2018-c-fcloud10-11}
with (without loss of generality
and for $0< \angle {\bf a},{\bf b} \le \frac{\pi}{4}$)
 ${\bf a} =\begin{pmatrix} 1,0,0 \end{pmatrix}$
and
${\bf b} =\frac{1}{\sqrt{1+x^2}}\begin{pmatrix} x,1,0 \end{pmatrix}$
 yields
the new terminal vector of the TITS/10-cloud
${\bf d} =\frac{1}{\sqrt{1+x^2}}\begin{pmatrix} 1,-x,0 \end{pmatrix} =  u_{20}$
which already occurs as the vector  $ u_{20}$
in the original TIFS/10-gadget.
The only additional vertex ${\bf c} =\begin{pmatrix} 0,0,1 \end{pmatrix}$
is from the edge connecting $u_{1}$ with $u_{3}$, as well as $u_{20}$ with $u_{22}$.
thereby ``completing'' the two cliques/contexts $\{u_1,c,u_3\}$ and  $\{u_{20},c,u_{22}\}$.
The angle between the two terminal points $u_1$ and $u_{20}$
of this TITS/11-gadget is $0 < \arccos \frac{1}{\sqrt{1+x^2}}\le \frac{\pi}{4}$ radians ($45^\circ$) as $0<x\le 1$.
This configuration is also a
TITS/10-cloud for the 17~pairs
$u_1-\{u_8,u_9,u_{12},u_{13},u_{16},u_{17},u_{22}\}$,
$u_{6}-u_{22}$,
$u_{7}-\{u_{12},u_{16},u_{22}\}$,
$u_{9}-u_{14}$,
$u_{10}-u_{22}$,
$u_{11}-\{ u_{16}, u_{22}\}$,
$u_{14}-u_{22}$, and
$u_{15}-u_{22}$, respectively.

\section{Discussion}

It is important to notice that, for {\em fixed terminal vertices}, depending on the {\em cloud chosen}, very different classical predictions follow.
Indeed, once the terminal vertices are fixed, it is not too difficult to enumerate a quantum cloud which,
interpreted classically, predicts and demands {\em any} kind of input-output behavior.
This renders an element of arbitrariness in the interpretation of quantum clouds.

The relevance of this observation lies in the conceivable interpretation of elementary empirical observations,
such as a single particular click in a detector.
Suppose a quantum is prepared in a pure state ``along'' a unit vector ${\bf a}$ and, when measured ``along''
$\textsf{\textbf{E}}_{\bf b}={\bf b}^\dagger  {\bf b}$,
``happens to activate a detector'' corresponding to that state ${\bf b}$;
that is, a detector associated with this latter property clicks.
Depending on the quantum cloud considered, the following contradictory claims are justified:
\begin{enumerate}
\item
if the quantum cloud allows both values then the claim is that there is no determination of the outcome; the event ``popped up'' from nowhere, {\it ex nihilo},
or, theologically speaking, has come about by {\it creatio continua} (cf. Kelly James Clark's God--as--Curler metaphor~\cite{Clark-2017-GodAsCurler});
\item
in the case of a 10-gadget the system is truly quantum and cannot be classical;
\item
in the case of an 11-gadget the  system could be classical;
\item
in case of a cloud inducing value indefiniteness the claim can be justified that the system cannot be classical, as no such event (not even its absence)
should be recorded. Indeed, relative to the assumptions made, the (non)occurrence of any event at all is in contradiction to the classical predictions.
\end{enumerate}
Conversely, if the experimenter observes no click in a detector associated with the  state $  {\bf b} $,
then, depending on the quantum cloud considered, the following contradictory claims are justified:
\begin{enumerate}
\item
as mentioned earlier, if the quantum cloud allows both values then there exists {\it creatio continua}
(currently, this appears to be the orthodox majority position);
\item
in the case of a 10-gadget the system could be classical;
\item
in case of an 11-gadget the system is truly quantum and cannot be classical;
\item
just as mentioned earlier, in case of a cloud inducing value indefiniteness the claim can be justified that the system cannot be classical, as no such event (not even its absence)
should be recorded.
\end{enumerate}

As a result, depending on the quantum cloud considered, any (non)occurrence of some single outcome can be published
(or rather marketed in venerable scientific journals)
as a crucial experiment indicating that the associated system cannot be classical.
Likewise, by taking other quantum clouds, any such outcome may be considered to be consistent with classicality:
(non)classicality turns out to be {\em means relative} with respect to the quantum clouds considered.
As quantum clouds are configurable for any input-output port setup this is true for any measurement outcome.

The situation turns even more precarious if one considers quantum clouds with a nonunital (and nonseparable) set of two-valued states,
such as the ones
depicted in Figs.~\ref{2018-c-f-TIFS} and~\ref{2018-c-f-TITS}:
In the particular faithful orthogonal
representation~\cite[Table~1, p.~102201-7]{2015-AnalyticKS}
the vector along
$\frac{1}{\sqrt{10}}\begin{pmatrix} 2\sqrt{2},1,-1 \end{pmatrix}$
yields a classical prediction amounting to the {\em nonoccurrence} of the particular
quantum observable. For another example
take a cloud introduced by Tkadlec~\cite[Fig.~2]{tkadlec-96}.
It is based on a set of orthogonal vectors
communicated to Specker by Sch\"utte~\cite{clavadetscher} and contains
36~vertices in 26~contexts/cliques which allow 6~two-valued states
enforcing 8~vertices to be $0$. In the particular faithful orthogonal representation of Tkadlec,
those correspond to the vectors along
$\begin{pmatrix} 1,0,0 \end{pmatrix}$,
$\begin{pmatrix} 0,0,1 \end{pmatrix}$,
$\begin{pmatrix} 1,0,1 \end{pmatrix}$,
$\begin{pmatrix} 1,0,-1 \end{pmatrix}$,
$\begin{pmatrix} 2,0,-1 \end{pmatrix}$,
$\begin{pmatrix} 1,0,2 \end{pmatrix}$,
$\begin{pmatrix} -1,0,2 \end{pmatrix}$, and
$\begin{pmatrix} 2,0,1 \end{pmatrix}$.
At the same time the vector
$\begin{pmatrix} 0,1,0 \end{pmatrix}$ is forced to be $1$.
Since without loss of generality, an orthogonal transformation can transform all of
these vectors into arbitrary other directions (while maintaining angles between vectors and,
in particular, orthogonality)
the assumption of such unital configurations
and their classical interpretation
immediately yields any desired contradiction
with any individual measurement outcome.

This arbitrariness could be overcome by some sort of ``superselection rule''
prioritizing or selecting particular quantum clouds over other ones.
However, in the absence of such superselection rules a generalized Jayne's principle, or rather Laplace's principle of indifference,
implies that any choice of a particular quantum cloud over other ones
amounts to an ``epistemic massaging'' of empirical data, and their nonoperational,
misleading overinterpretation in terms of a speculative ontology~\cite{berkeley,stace,Goldschmidt2017-idealism};
or, to quote Peres~\cite{peres222}, {\em unperformed experiments have no results''.}
In contradistinction, it may not be too speculative to hold it for granted that the only operationally
justified ontology is the assumption of a single one context or its associated maximal observable.

\begin{acknowledgments}
The author acknowledges the support by the Austrian Science Fund (FWF): project I 4579-N and the Czech Science Foundation (GA\v CR): project 20-09869L.

The author declares no conflict of interest.

I kindly acknowledge enlightening discussions with Adan Cabello, Jos\'{e} R. Portillo, and Mohammad Hadi Shekarriz.
I am grateful to Josef Tkadlec for providing a {\em Pascal} program
which computes and analyses the set of two-valued states of collections of contexts.
All misconceptions and errors are mine.
\end{acknowledgments}








\end{document}